\documentclass[prb,twocolumn,superscriptaddress,showpacs,amsmath,amssymb]{revtex4-2}

\usepackage{graphicx}
\usepackage{latexsym}
\usepackage{amsmath}
\usepackage{amssymb}
\usepackage{amsfonts}
\usepackage{color}
\usepackage{bm}
\usepackage{verbatim}
\usepackage{float}

\definecolor{MS-color}{RGB}{255,0,0}

\usepackage{framed}
\definecolor{shadecolor}{RGB}{222,222,221}
\begin{document}
 
  

\title{Ultrastrong magnon-photon coupling and entanglement in superconductor/ferromagnet nanostructures }

\date{\today}

 \author{Mikhail Silaev}
\affiliation{Computational Physics Laboratory, Physics Unit, Faculty of Engineering and Natural Sciences, Tampere University, P.O. Box 692, Tampere, Finland}

 \begin{abstract}
 Ultrastrong light-matter coupling opens exciting possibilities to generate squeezed quantum states and entanglement. 
  Here we propose a way to achieve this regime  in superconducting hybrid nanostructures with ferromagnetic interlayers.
 Strong confinement of electromagnetic field between superconducting plates is found to result in the existence of 
   magnon-polariton modes with ultrastrong magnon-photon coupling, ultra-high cooperativity and very large group velocities.
  These modes provide a numerically accurate explanation of recent experiments and have intriguing quantum properties. 
   { The magnon-polariton quantum vacuum consists of the squeezed magnon and photon states with the degree of
    squeezing controlled in wide limits by the external magnetic field. The ground state population of virtual photons 
    and magnons is shown to be very large which can be used for 
    generating correlated magnon and photon pairs.  Excited states of magnon-polaritons contain bipartite entanglement between magnons and photons.   This property can be used for transferring entanglement between different types of quantum systems.    }
   \end{abstract}

 \maketitle

 Cavity-enhanced light-matter interaction has become one of the most perspective tools to 
 control and study the properties of quantum materials  \cite{raimond2001manipulating,frisk2019ultrastrong, basov2017towards, bloch2022strongly,schlawin2022cavity}. In the core of this approach is the formation of hybrid quantum states consisting of the  the matter and electromagnetic field components. Especially pronounced such hybridization becomes in the strong-coupling regime when the coupling strength is larger than  decay rates of both the cavity and the quantum system states \cite{kaluzny1983observation,meschede1985one,thompson1992observation,lodahl2015interfacing,gu2017microwave}.
 %
In many-body quantum this regime leads to the formation of hybrid polaritons \cite{basov2016polaritons, basov2017towards} combining photons and various collective modes. Among them are the exciton-polaritons\cite{deng2010exciton}, 
magnon-polaritons (MP) \citep{lachance2019hybrid, rameshti2022cavity}
and hybrid superconducting modes  \citep{allocca2019cavity,curtis2019cavity,curtis2022cavity} .

 \begin{figure}[htb!]
 \centerline{
 $ \begin{array}{c}
 \includegraphics[width=0.5\linewidth] {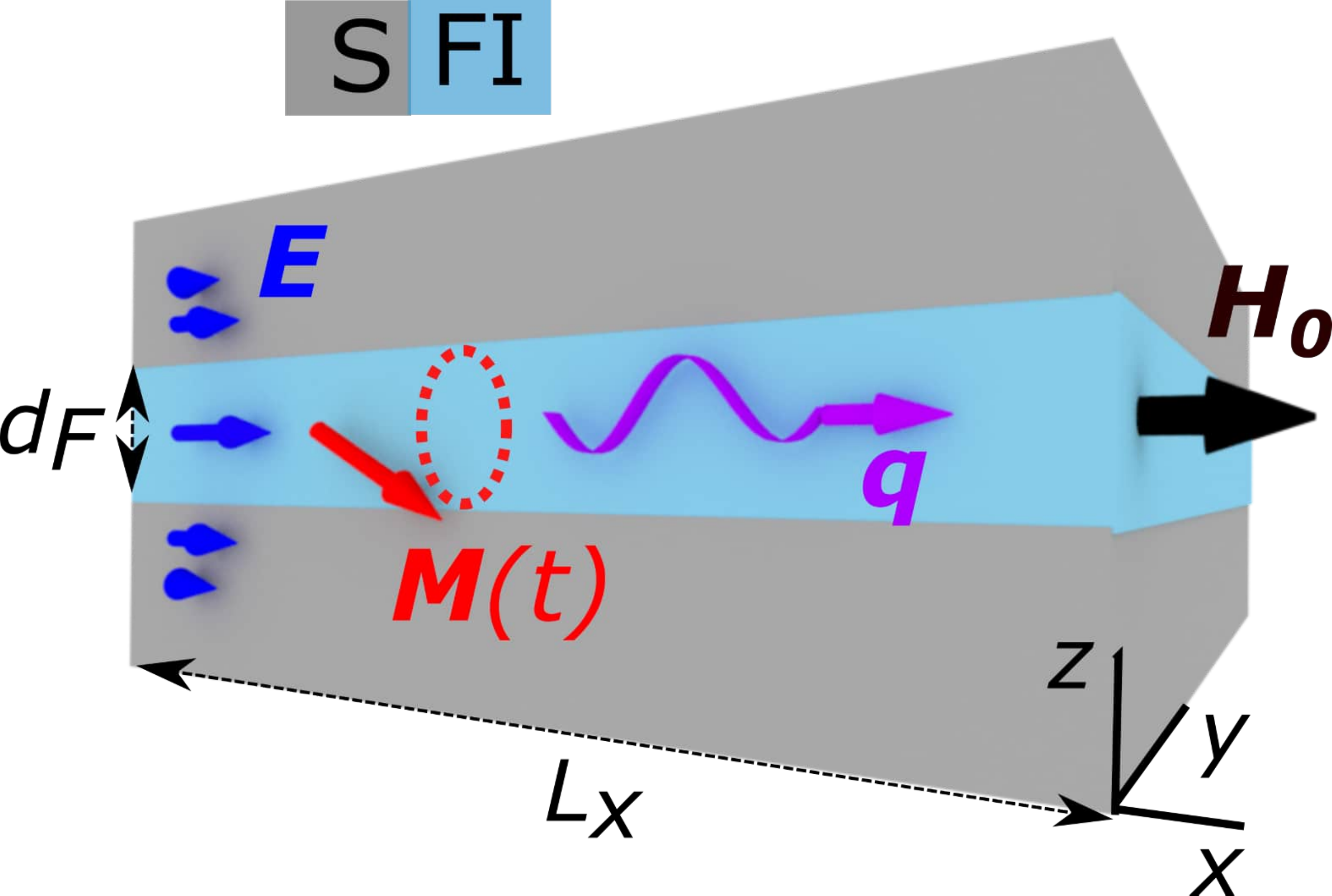}
 \put (-60,92) {\large (a) }
  \includegraphics[width=0.5\linewidth] {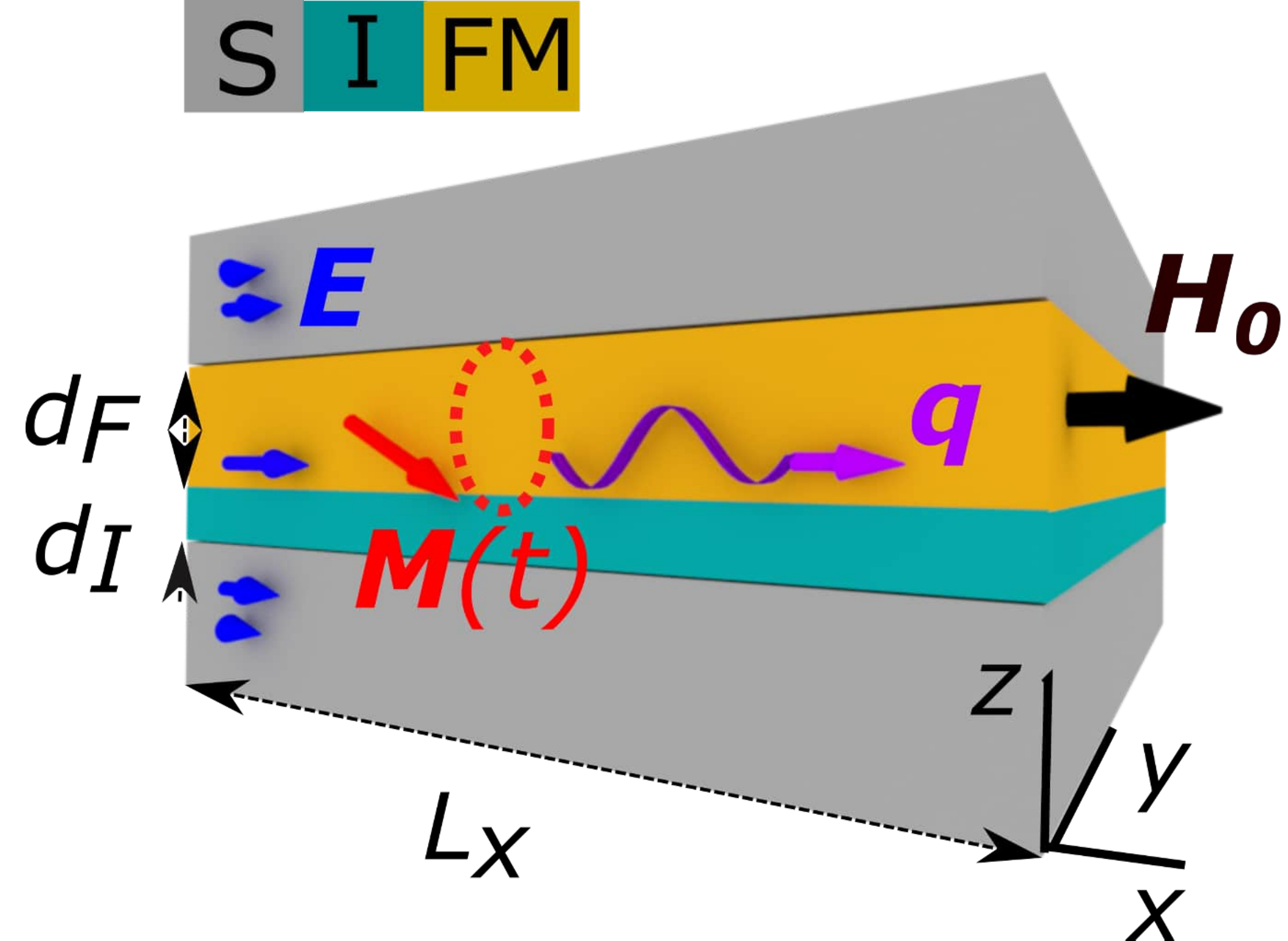}
 \put (-60,92) { \large (b) }
  \\
 \includegraphics[width=0.50\linewidth] {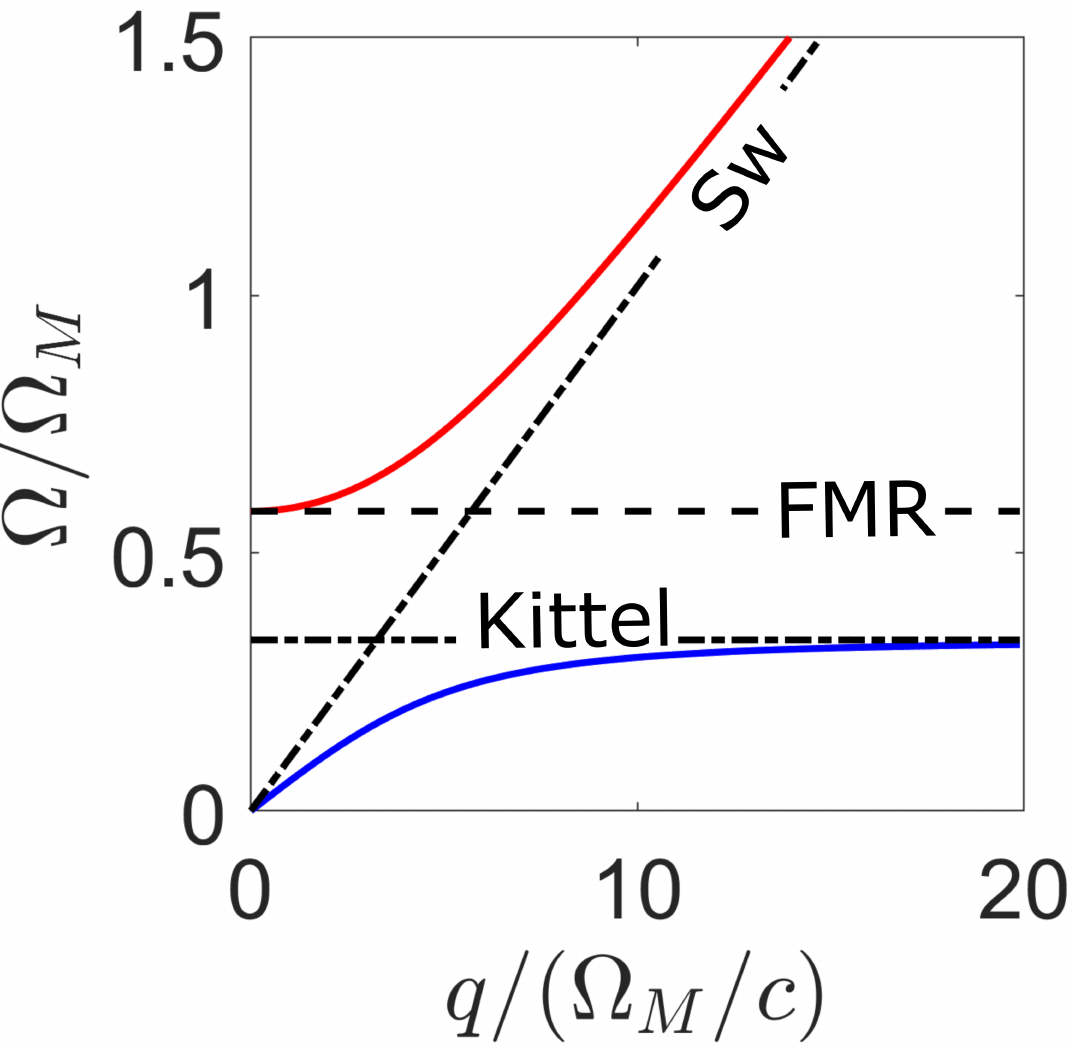}
 \put (-60,121) { \large (c) }
 \put (-85,100) {\color{red} \large $\Omega_{UP}$ }
 \put (-30,30) {\color{blue} \large $\Omega_{LP}$ }
 \;\;
 \includegraphics[width=0.50\linewidth] {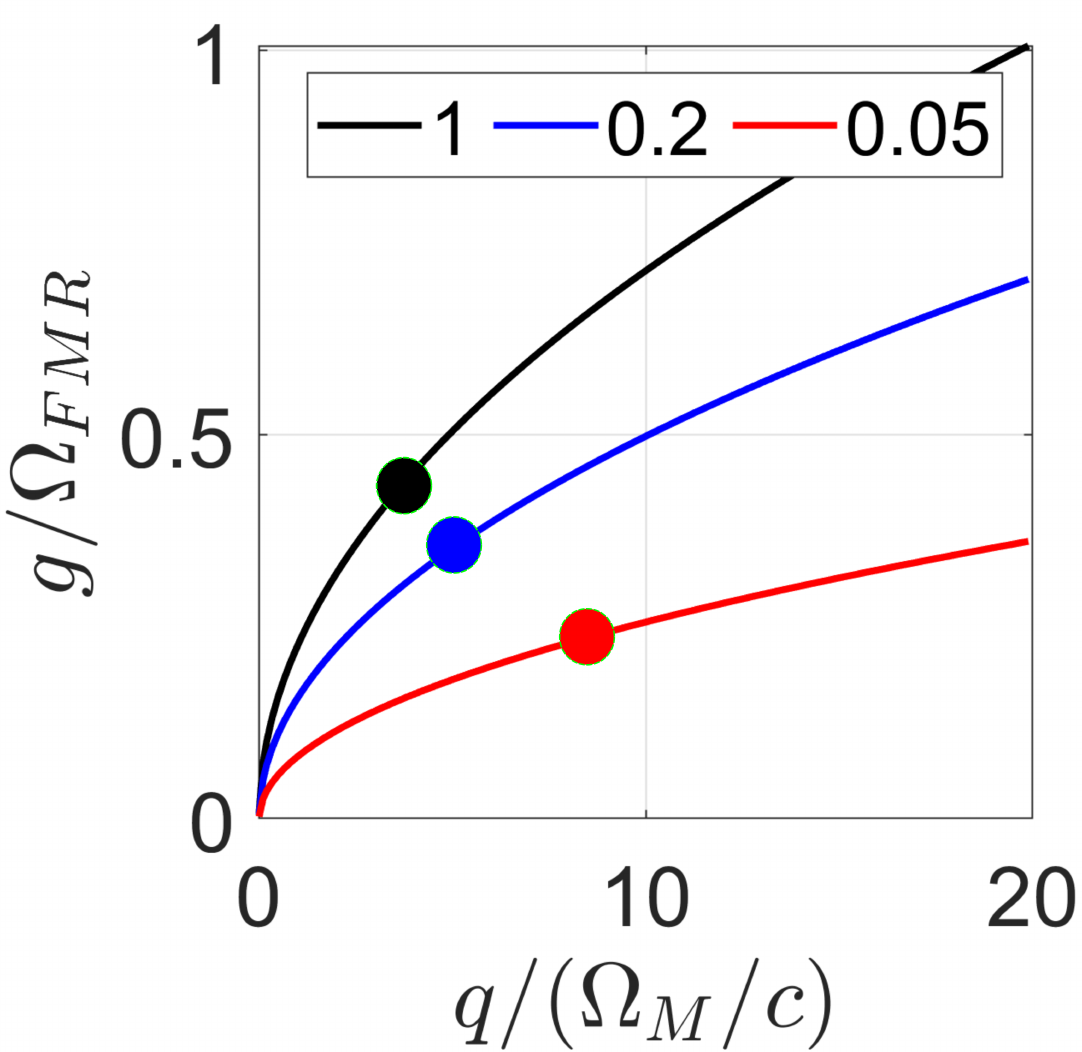} 
 \put (-60,121) {\large (d) }  
 \end{array}$
 }
 \caption{  \label{Fig:SystemSketch0} 
 (a,b) Superconducting heterostructures  hosting MP modes. (a) With ferro- or ferrimagnetic insulator (FI) interlayer.  (b) With 
 composite interlayer consisting of metallic ferromagnet (FM) and usual insulator (I).   The static external field is $\bm H_0= H_0 \bm x$, the precessing magnetization is $\bm M (t)$ and the wave vector of MP is $\bm q = q \bm x$. 
  %
 (c)MP spectrum in S/FI/S system with $d_{FI}=0.3\lambda$. Upper $\Omega_{UP}$ and lower $\Omega_{LP}$ MPs are shown by red and blue lines, respectively. Dashed line is FMR frequency $\Omega_{FMR}$, dotted line is a Swihart (Sw) mode frequency $\Omega_{Sw}$, 
 dashed-dotted is the Kittel frequency $\Omega_K = \gamma \sqrt{H_0B_0}$. 
 Parameters are $d_{F}/\lambda=0.5$, $H_0= 4\pi M_0/10$, $\Omega_M = 4\pi \gamma M_0$. 
 (d) Coupling parameter $g (q)$ Eq.(\ref{Eq:gGen}) for $d_{F}/\lambda = 1;\; 0.2;\; 0.05$ and 
 $H_0= 4\pi M_0/10$. Filled circles show  $g(q_{res})$. }
 \end{figure}
 
Even more exciting is an ultrastrong coupling regime when the 
interaction is comparable with the eigen frequencies of interacting modes
\cite{frisk2019ultrastrong,forn2019ultrastrong}.  
Many interesting effects based on the  hybridization between states 
with different number of excitations have been predicted to be observable in this realm.    
In view of the rapidly developing  quantum cavity magnonics \cite{lachance2019hybrid,rameshti2022cavity} 
it is very appealing to realize the ultrastrong-coupling regime in such systems where  "matter" is represented by the
magnons and the "light" is represented by the microwave cavity fields. 
It has been achieved in specially designed 3D microwave cavities \cite{kostylev2016superstrong} and 
in recently discovered on-chip superconducting nanostructures \cite{golovchanskiy2021approaching,golovchanskiy2019ferromagnet} 
combining superconducting (S) plates separated by the insulating (I)  \cite{swihart1961field} 
and  ferromagnetic metal (FM) interlayers  typically of $10 -100$ nm thickness. 
The ultrastrong photon-magnon coupling observed in S/FM/I/S \cite{golovchanskiy2021approaching} and S/FM/S/I/S \cite{golovchanskiy2019ferromagnet,golovchanskiy2021ultrastrong} systems is by several orders of magnitude larger than in 
S/FM\cite{li2019strong,hou2019strong} and S/FI bilayers \cite{huebl2013high}, where FI stands for the ferro- or ferrimagnetic insulator 
like yttrium iron garnet  (YIG)\cite{chumak2015magnon,serga2010yig}. 

In the present Letter we explain these experiments by developing the theory of 
magnon-polariton (MP) states for both the generic S/FI/S system and the experimentally studied more complex S/FM/I/S one with in-plane
stationary magnetization. Hybridization of spin and Josephson plasma waves has been suggested previously for S/FM/I/S systems with  perpendicularly magnetized FM films \cite{volkov2009hybridization}.
Our theory is very accurate to explain experiments \cite{golovchanskiy2021approaching}  without using adjusting 
parameters. Besides that we find several unique properties of MPs in such systems making them a versatile platform for classical and quantum magnonics \cite{chumak2015magnon,chumak2021roadmap,chumak2022roadmap}.

Let us consider the generic S/FI/S and S/FM/I/S systems  shown in Fig.\ref{Fig:SystemSketch0}a,b  
 hosting both photonic modes and magnons. The former is represented by highly confined electromagnetic field solutions 
 found by Swihart \cite{swihart1961field}. This Swihart mode is localized within the layer of the thickness $d_{F} + 2\lambda$ and  $d_{I} + d_F + 2\lambda$ in S/FI/S and S/FM/I/S systems, respectively, where $\lambda$ is the London penetration length.   For typical superconducting material Nb\cite{golovchanskiy2021approaching,golovchanskiy2021ultrastrong} and nanostructure parameters $d_{F}, d_I \sim \lambda $ the field  
is localized within the layer much thinner than the photon wavelength which is $\sim 1-10$ mm  for 
typical microwave frequencies $\omega \sim 10 -100 $ GHz. 
This strong confinement leads to the unusual polarization structure with electric field $\bm E$ almost 
 $\parallel \bm q$ as shown in Figs.\ref{Fig:SystemSketch0}a,b by blue arrows. Simultanoesly it leads to the 
 strongly enhances magnon-photon interaction as compared to the 3D cavities\cite{lachance2019hybrid,rameshti2022cavity}.
 Dispersion of the Swihart mode is
 \begin{align} \label{Eq:SwSFIS}
& {\rm in\; S/FI/S:}  & \Omega_{Sw} (q)   =  
 cq \sqrt{ d_{F}/\varepsilon( d_{F} + 2\lambda) }
 \\  \label{Eq:SwSFMIS}
& {\rm in\; S/FM/I/S:} & \Omega_{Sw} (q)   =  
 cq \sqrt{ d_I/\varepsilon(d_I + d_{F} +2\lambda ) }
 \end{align} 
 where $c$ is light velocity and $\varepsilon \sim 10$ is the dielectric constant in  YIG 
 or usual insulators  Al$_2$O$_3$, Si. 
 
 Magnons in S/FI/S and S/FM/I/S systems are excitations of the  magnetization direction $\bm M (t)$ in the magnetic layer. 
  In thin films with\cite{schneider2021control, golovchanskiy2021approaching} $d_{F}\sim 100$ nm  
 the scale of magnon frequency dispersion \citep{damon1961magnetostatic,kalinikos1980excitation,prabhakar2009spin} 
   $\sim d_{F}^{-1}$ is much larger than  the wavenumbers of microwave photons $q \sim 0.1-1$ mm$^{-1}$. Therefore magnon frequency can be assumed constant coinciding with the fundamental ferromagnetic resonance (FMR) mode $\Omega_{FMR}$. In this regime magnons play the role analogous to the  electronic atomic  \cite{rempe1987observation} or intersubband transition in cavity electrodynamics \cite{ciuti2005quantum}. The cavity is represented by the Swihart mode Eq.(\ref{Eq:SwSFIS},\ref{Eq:SwSFMIS}). The resonance when $\Omega_{Sw}(q_{res}) = \Omega_{FMR}$ corresponds to the wavelengths $q^{-1}_{res} \sim 1 - 10$ mm. The anticrossing between magnon and Swihart modes shown in Fig.\ref{Fig:SystemSketch0}c results in two magnon-polariton (MP) modes. Detailed calculations yielding Fig.\ref{Fig:SystemSketch0}c are presented below.

{
 Our starting equations for the frequency components
 of magnetization $\bm M_\omega$, magnetic field
 $\bm H_\omega$ and induction $\bm B_\omega = \bm H_\omega + 4\pi \bm M_\omega$ read 
 \begin{align} \label{Eq:LLG}
 & i\omega {\bm M}_\omega = \gamma (\bm B_0\times\bm M_\omega + \bm  
 B_\omega \times\bm M_0  )
  \\ \label{Eq:HomegaMaxwellWave}
 & \bm\nabla\times ( {\tilde \varepsilon}^{-1} \bm\nabla\times \bm H_\omega ) - q_v^2 \bm B_\omega =0
  \end{align} 
Here Eq.(\ref{Eq:LLG}) is Landau-Lifshitz-Gilbert (LLG) one,  $\bm B_0 = \bm H_0 + 4\pi \bm M_0$ is the stationary magnetic field. 
The gradient terms are neglected since length scales are much larger than the exchange length. 
Maxwell equations result in Eq.(\ref{Eq:HomegaMaxwellWave})  where $q_v= \omega/c$ is the wave number in vacuum. 
 In the insulator, either  FI or I we have  $\tilde{\varepsilon} = \varepsilon$
 is the dielectric constant while in metal  $\tilde{\varepsilon} = - 4\pi i \sigma/\omega$. 
 The conductivity $\sigma$ in FM is $\sigma_F = const$ while in S   $ \sigma_S (\omega) =  c^2/(4\pi i \lambda^2 \omega)$,
 where $\lambda$ is the London penetration length. 

First consider S/FI/S and S/FM/I/S systems shown in Figs.\ref{Fig:SystemSketch0}a,b
with $L_x=\infty$ so that all fields perturbations $\propto e^{i q x}$ with arbitrary $q$. Time-dependent magnetization components are $\bm M_\omega = (0, M_y, M_z)$. Due to the presence of metallic S layers we can simplify the problem in the long-wavelength limit\cite{braude2004excitation}  $q \lambda \ll 1$. In this case Eq.(\ref{Eq:HomegaMaxwellWave}) yields $B_z=0$ and $H_x=0$ in metallic S. Due to the continuity of $B_z$ and $H_x$ they are small also in the attached I and FI layers. Then we are left with equations only for the $H_y$ component in each layer. 
 Boundary conditions at interfaces  follow directly from Maxwell equations yielding  the continuity of tangential components 
 $H_y$ and $E_x$, where $E_x=(i/\tilde \varepsilon q_v)\nabla_z H_y  $ .
 
{
  First we consider S/FI/S system shown in Fig.\ref{Fig:SystemSketch0}a. 
 Solving equations for $H_y$ in S and FI layers and matching them with the help of above boundary conditions
  \cite{SupplMat}  we get the equation for 
  MP dispersion relation 
  \begin{align} \label{Eq:DickeDispersion}
 (\omega^2- \Omega_{Sw}^2)(\omega^2 -\Omega_{FMR}^2) = 4 g^2 
 \Omega_{FMR} \Omega_{Sw} 
 \end{align}
 Here $\Omega_{Sw} (q)$ is the Swihart mode frequency Eq.(\ref{Eq:SwSFIS}), 
  $\Omega_{FMR}$ is the FMR frequency in S/FI/S system given by  
  \begin{align} 
  \label{Eq:OmegaFMRFI}
  & \Omega_{FMR}  =  
 \gamma\sqrt{H_0B_0 + \frac{4\pi M_0B_0 d_{F}}{  d_{F} + 2\lambda  }  } 
   \end{align}   
   where the first term in r.h.s. $\Omega_K = \gamma \sqrt{H_0B_0}$ is the FMR frequency in isolated ferromagnetic film 
   derived by Kittel \cite{kittel1948theory}.
The coupling parameter equivalent to the vacuum Rabi splitting \cite{rempe1987observation, weisbuch1992observation, raimond2001manipulating, ciuti2005quantum} is given by
 \begin{align} \label{Eq:gGen}
  g= \frac{1}{2}\sqrt{ \frac{\Omega_{Sw} }{\Omega_{FMR}} } \sqrt{\Omega_{FMR}^2 - \Omega_K^2}
 \end{align}  
 
  { Now let us consider the S/FM/I/S system Fig.\ref{Fig:SystemSketch0}b studied in recent experiment 
  \citep{golovchanskiy2021approaching}.  This system also hosts MPs  \cite{SupplMat} determined by  
  Eq.(\ref{Eq:DickeDispersion}) with Swihart mode Eq.(\ref{Eq:SwSFMIS}) and FMR frequency
 \begin{align} 
 \label{Eq:OmegaFMRFM}
 & \Omega_{FMR}  =  
 \gamma\sqrt{H_0B_0 + \frac{4\pi M_0 B_0 d_{F} }{d_I + d_{F} + 2\lambda} }
 \end{align}  
 The coupling parameter $g$  is given again by Eq.(\ref{Eq:gGen}).
        
Solution of Eq.(\ref{Eq:DickeDispersion}) consists of upper and lower magnon-polariton modes with  frequencies $\Omega_{UP}(q)$ and
 $\Omega_{LP} (q)$, respectively, where $\Omega_{UP} > \Omega_{LP} $. 
 The example of dispersion curves for S/FI/S system with $d_{F}=0.5 \lambda$ are shown in Fig.\ref{Fig:SystemSketch0}c.   
 The coupling parameter Eq.(\ref{Eq:gGen}) is determined by the detuning  $ g \propto \sqrt{ \Omega_{FMR}^2- \Omega_{K}^2}$
 which strongly depends on thickness $d_{F}$ and external parameters such as 
 $H_0$ and temperature $T$ through the London length $\lambda (T)$.
Shown in Fig.\ref{Fig:SystemSketch0}b are the dependencies $g(q)$ for S/FI/S
 for $d_{F}/\lambda =1; 0.2; 0.05$.  As one can see it is possible to achieve an ultrastrong photon-magnon coupling 
$g \sim \Omega_{FMR}$. To demonstrate explicitly this consider a resonant point\cite{ciuti2005quantum} $q_{res}$
where $\Omega_{Sw} (q_{res}) = \Omega_{FMR}$. 
The coupling parameter at resonance in S/FI/S system is 
$g(q_{res}) = \gamma \sqrt{\pi d_{F} B_0M_0 /(  d_{F} + 2\lambda)  }  $.
 In the limit $H_0=0$ it can reach the maximal value of $ g(q_{res}) =\Omega_{FMR}/2$
 corresponding to the ultra-strong coupling regime\cite{frisk2019ultrastrong,ciuti2005quantum}. 
 Since there is an upper boundary on coupling $g(q_{res})< \Omega_{FMR}/2$ it is not possible to enter the deep-strong-coupling\cite{casanova2010deep,  bayer2017terahertz,yoshihara2017superconducting} regime and the super-radiant transition.  
    Similar result is valid for the S/FM/I/S system. 
 
 {\color{black} The dispersion relation Eq.(\ref{Eq:DickeDispersion}) can be written in the form introduced to fit experimental data
  \cite{golovchanskiy2021approaching}, which follows from the Hopfield Hamiltonian\cite{bayer2017terahertz, frisk2019ultrastrong,baranov2020ultrastrong,mueller2020deep}   
 \begin{align} \label{Eq:OmegaDispGol}
 \omega^4  - \omega^2 (\Omega_{Sw}^2 + \Omega_K^2 + 4 \tilde{g}^2) +  \Omega_{Sw} \Omega_K =0
 \end{align}
  with the renormalized coupling $\tilde{g} = \sqrt{\Omega_{FMR}^2 - \Omega_K^2} $. 
  It can be written as 
  \begin{align}
 {\rm in\; S/FI/S:\;} & \tilde g = \gamma \sqrt{\pi M_0B_0 d_{F} /(d_{F} + 2\lambda)}
   \\
  {\rm in\; S/FM/I/S:\;} & \tilde g = \gamma \sqrt{\pi M_0B_0 d_{F} /(d_{F} + d_I + 2\lambda)} 
  \end{align}
   Note that for small external fields $H_0\ll 4\pi M_0$ the coupling coefficient is almost constant 
 at  $\tilde g(H_0) \approx const $. As shown in Ref.\cite{golovchanskiy2021approaching} this allows obtaining accurate
  fits of experimental data using Eq.(\ref{Eq:OmegaDispGol})
 with $\tilde g$ as a phenomenological adjusting parameter. The present theory explains experiment quantitatively
  without using any adjusting parameters. 
  }
 
Both S/FI/S and S/FM/I/S systems provide ultra-high cooperativity for MP states.  
The cooperativity is $C = g(q_{res}) / (\alpha_{ph}\alpha_{mag})$, where  the photon and magnon decay rates are  $\alpha_{ph}$ and
 $\alpha_{mag}$, respectively. 
For S/FI/S system with YIG we assume \cite{huebl2013high}  $\alpha_{ph} =3$ MHz, $\alpha_{mag} = 50$ MHz and
 $\Omega_{M} \approx 6$ GHz. Then for the parameters corresponding to Fig.\ref{Fig:SystemSketch0}c
we get from Eqs.(\ref{Eq:OmegaFMRFI},\ref{Eq:gGen}) $C = 1.3\; 10^4$.

Similar estimation can be made for S/FM/I/S system with Py ferromagnet and parameters corresponding to experiment \cite{golovchanskiy2021approaching} $d_I =d_{F} = 0.3 \lambda$.
Taking the decay rates \cite{hou2019strong} $\alpha_{ph} =0.7$  MHz,  $\alpha_{mag} = 200$ MHz, 
 $\Omega_M = 31.3$ GHz and $H_0 = 4\pi M_0/10$, from 
 Eqs.(\ref{Eq:gGen},\ref{Eq:OmegaFMRFM}) we get much larger cooperativity  $C = 3.7\; 10^5 $. This value is 
 comparable with the ultra-high mangon-photon cooperativity obtained in specially designed 3D cavities with 
 with focused magnetic fields \citep{goryachev2014high}. 
 
  In the region  of strong magnon-photon mixing $q\approx q_{res}$ in Fig.\ref{Fig:SystemSketch0}c the group velocities of MP branches 
  $v_{j} = \partial \Omega_{j}/\partial q $ are of the order $v_{j} \sim c/20 \approx 1.5 \; 10^4$ km/s.  This velocity is $10^3$ times larger than that  of the fastest known magnons \cite{chen2017group}. In complement to magnon gating by narrow S stripes\cite{yu2022efficient} 
  and magnon-condensate coupling \cite{johnsen2021magnon} the present 
 S/FI/S and S/FM/I/S systems are extremely efficient in transmitting magnetic signals which is promising for the ultra-fast and energy-efficient data processing  \cite{chumak2021roadmap}.  
  
  In several limiting cases MPs feature interesting behaviour.  
{\bf (i)} For $H_0=0$ Eq.(\ref{Eq:gGen}) yields  $g=\sqrt{\Omega_{Sw}\Omega_{FMR}}/2$ so that 
$\Omega_{LP}(H_0=0)=0$ and $\Omega_{UP}(H_0=0) = \sqrt{\Omega_{Sw}^2 +\Omega_{FMR}^2} $. 
 The asymptotic behaviour for small fields $H_0\ll M_0$ is $\Omega_{LP}= \Omega_{K} \Omega_{Sw}/\sqrt{\Omega_{Sw}^2 +\Omega_{FMR}^2} $. 
 As shown below, this behaviour is crucial for realizing highly squeezed vacuum magnon and photon states. 
 {\bf (ii)} For large wave numbers $q \gg \Omega_M/c$ the asymptotic is  $\Omega_{UP} (q\to \infty)= \Omega_{Sw}$ 
 and $\Omega_{LP}(q\to \infty)=\Omega_K$ as can be seen in Fig.\ref{Fig:SystemSketch0}c.  
This behaviour explains earlier theoretical result \cite{2207.13201} and experiments\cite{li2018possible}
 featuring no shift of measured FMR frequency from  $\Omega_{K}$ in S/FM/I/S systems  much shorter than the wavelength  $L_x \ll c/\Omega_M$ .
 In such systems it is not the genuine FMR frequency $\Omega_{FMR}$ Eq.(\ref{Eq:OmegaFMRFM}) which is measured but the 
 that of lower MP modes having $\Omega_{LP}(q_n) \approx \Omega_K $ for all standing waves with quantized wavenumbers 
 $q_n= \pi n/L_x$ for integer $n$.  
  {\bf (iii)} For $d_I\to 0$  the Swihart mode velocity in S/FM/I/S system becomes very small so that according to Fig.\ref{Fig:SystemSketch0}c the lower MP mode disappears $\Omega_{LP} (q) \to 0$. At the same time the upper MP mode becomes dispersion-less $\Omega_{UP}(q) \approx \Omega_{FMR}$ given by Eq.(\ref{Eq:OmegaFMRFM}) with $d_I=0$. This result coincides with previous calculations\cite{2207.13201} and shows that experiments \cite{li2018possible, jeon2019effect,golovchanskiy2020magnetization} in S/FM/S systems  have measured in fact the 
  upper MP modes having $\Omega_{UP}(q_n) \approx \Omega_{FMR} $ for all integer $n$.

 {  The quantization of MP states has been observed in recent experiments \cite{golovchanskiy2021approaching,golovchanskiy2021ultrastrong}.
 To explain them note that in general structures of finite length $L_x$ shown in Fig.\ref{Fig:SystemSketch0}a,b host standing waves of MPs 
 with discrete momenta $q_n=\pi n /L_x$ with integer $n$  corresponding to the quantized MP states.
  Their discrete frequencies $\Omega_{LP,n}$ and $\Omega_{UP,n}$
  are given directly by 
 Eq.(\ref{Eq:DickeDispersion}) with $q=q_n$, that is $\Omega_{j,n} = \Omega_{j}(q_n)$ where $j = \{ LP, UP \}$. 
 Let us consider  S/FM/I/S structure shown in Fig.\ref{Fig:SystemSketch0}b 
 with parameters precisely those used in experiment \cite{golovchanskiy2021approaching}
 $\lambda= 80$ nm, $d_F= 25$ nm, $d_I =13$ nm,  $L_x = 1.1$ mm, $\varepsilon =10$, $M_0 =1.06$ T so that $\Omega_M = 31.3$ GHz.
 In Fig.\ref{Fig:DispersionMPH0}a,b 
 the blue and red lines show $\Omega_{LP,n}( H_0)$ and $\Omega_{UP,n}(H_0)$, respectively for $n=1,2,3$.
 As shown in Fig.\ref{Fig:DispersionMPH0} the sequence of modes $\Omega_{LP,n}( H_0)$ is bounded from above by 
 the Kittel frequency since as discussed above  $\Omega_{LP} (q\to \infty) = \Omega_K$. 
 The other sequence $\Omega_{UP,n}( H_0)$ grows unbounded with $n$. 
 To compare with experiment \cite{golovchanskiy2021approaching} in Fig.\ref{Fig:DispersionMPH0}b we zoom in the  domain which 
 is marked by the green rectangle in Fig.\ref{Fig:DispersionMPH0}a. 
 There is a very accurate agreement between experimental data \cite{golovchanskiy2021approaching} 
 shown by open circles and theory curves shown by solid lines. Note that we use no adjusting parameters.  
 Qualitatively similar results are expected for S/FI/S system in Fig.\ref{Fig:SystemSketch0}b. 
 \begin{figure}[htb!]
 \centerline{ 
 $ \begin{array}{c}
 \includegraphics[width=1.0\linewidth] {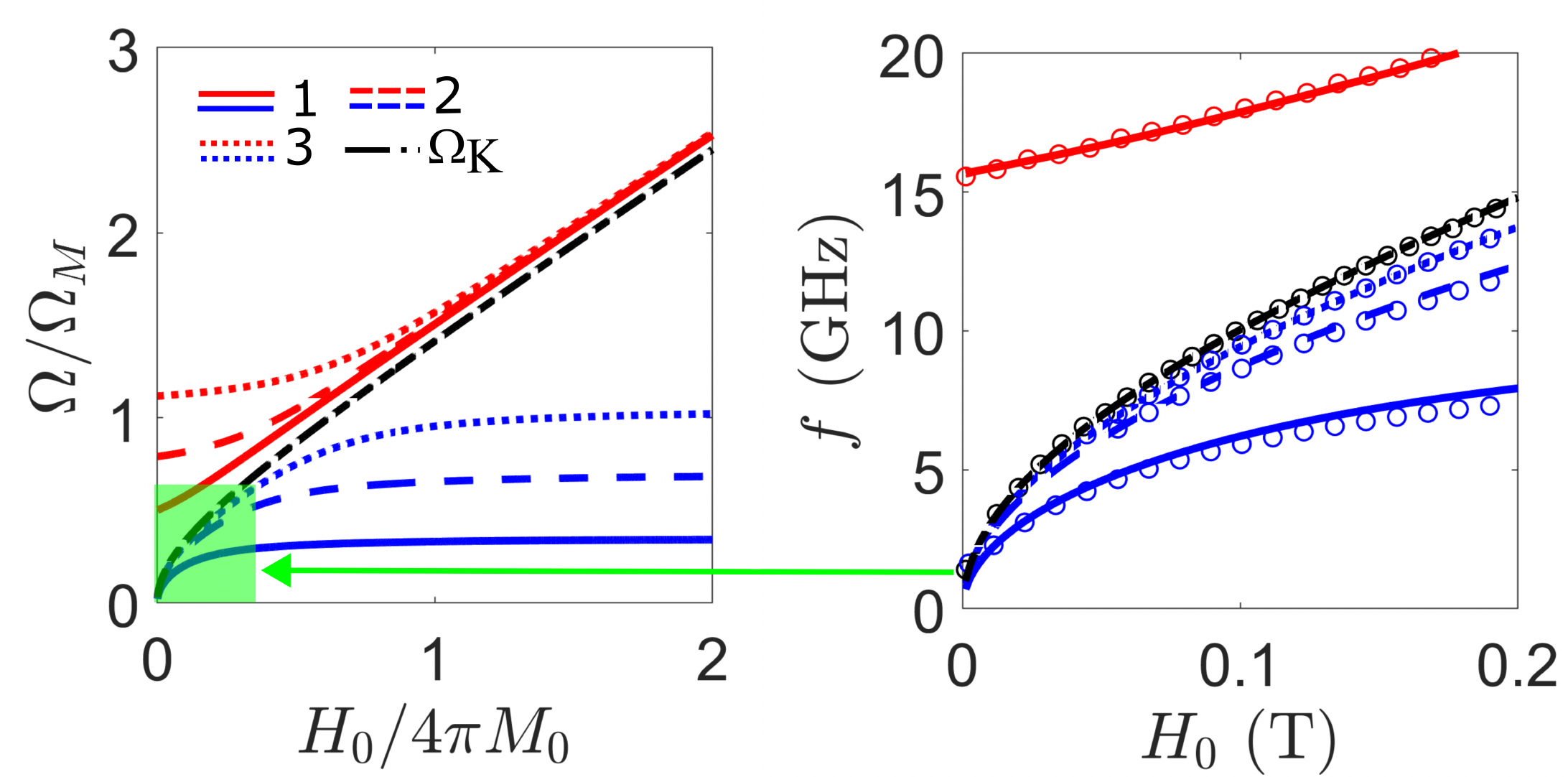}
 \put (-85,80) {\color{red}  $\Omega_{UP}$ }
 \put (-60,35) {\color{blue}  $\Omega_{LP}$ }
 \put (-190,125) { \large (a) }
 \put (-65,125) { \large (b) }
  \end{array}$  }
 \caption{\label{Fig:DispersionMPH0}
 (a) Quantized MP modes in S/FM/I/S system with parameters as in experiment \cite{golovchanskiy2021approaching},
 see text for details.  Upper (red lines) $\Omega_{UP, n}(H_0)$ and lower (blue lines)
  $\Omega_{LP, n}(H_0)$ modes with $n=1$ (solid lines), $n=2$ (dashedlines), $n=3$ (dotted lines). 
  The black dash-dotted line is the Kittel frequency $\Omega_{K} (H_0)$. 
   (b) The zoomed-in region  marked by green rectangle in panel (a) plotted in real units. 
   Open circles show experimental data  \cite{golovchanskiy2021approaching}.
 }
 \end{figure}
 
 \begin{figure}[htb!]
 \centerline{ 
 $ \begin{array}{c}
 \includegraphics[width=1\linewidth] {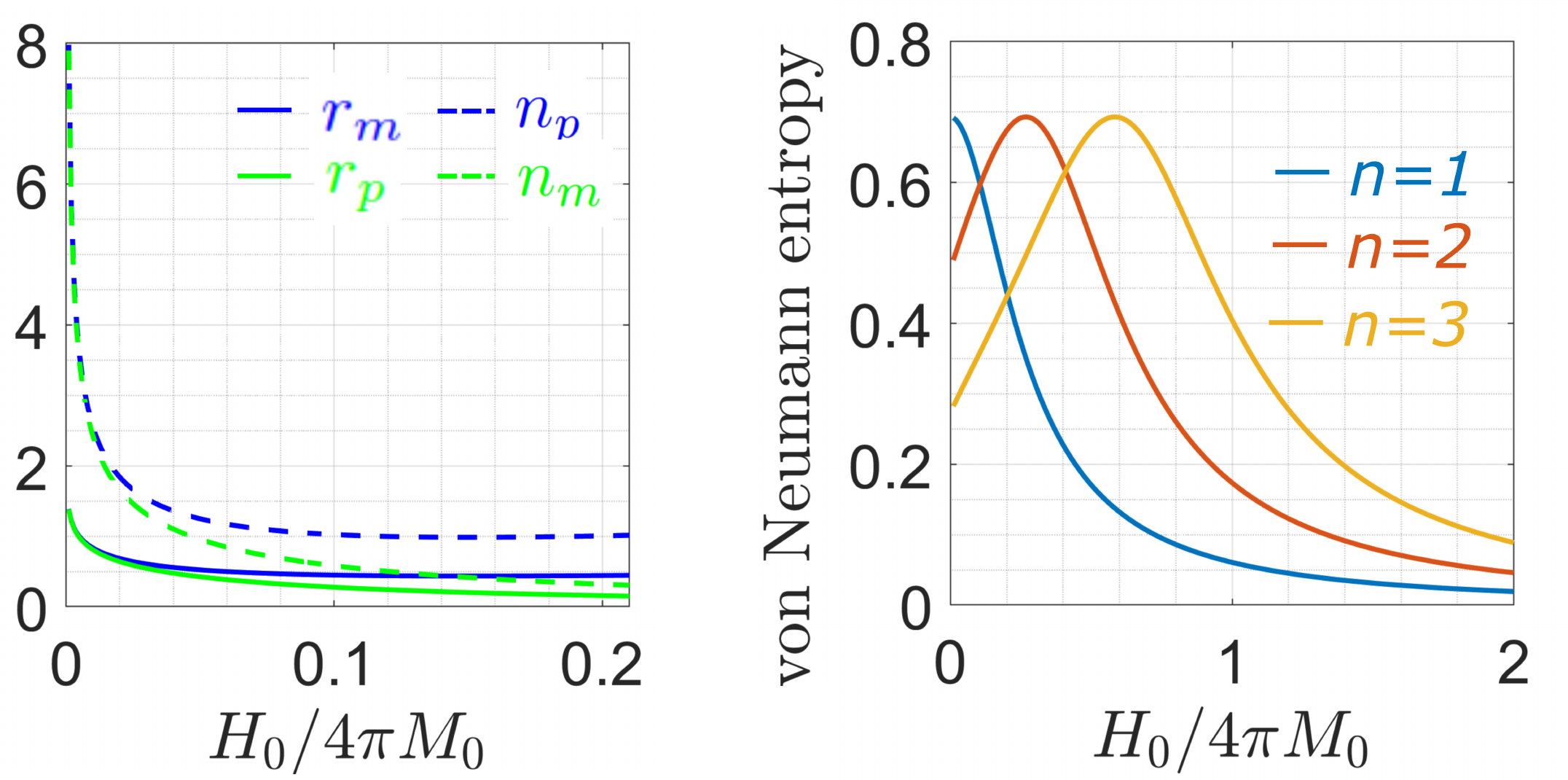}
  \put (-190,125) { \large (a) }
 \put (-65,125) { \large (b) }
  \end{array}$  }
 \caption{\label{Fig:FractionsH0}
 (a) Magnon $r_m$ and photon $r_p$ squeeze parameters for the vacuum state of lower MP mode with $n=1$ in in S/FM/I/S system.
 (b) von Neumann entropy measuring the photon-magnon entanglement in the excited states of lower MP mode with  $n=1;\; 2;\; 3$ in S/FM/I/S system. Parameters are the same as in Fig.\ref{Fig:DispersionMPH0}.  }
 \end{figure}

 { 
  The hybridization of magnon and photon modes given by the dispersion relation (\ref{Eq:DickeDispersion}) is equivalent to that 
  given by the Dicke model Hamiltonian \cite {dicke1954coherence,hopfield1958theory, hepp1973superradiant, emary2003chaos, ciuti2005quantum}.
 \begin{align}  \label{Eq:DickeModelHamQuantized}
 & \hat H = \Omega_{Sw}  \hat a^\dagger \hat a  + 
 \Omega_{FMR} \hat b^\dagger \hat b   +
  g (\hat a  +  \hat a^\dagger ) 
  (\hat b^\dagger + \hat b )
 \end{align}
 where $\hat a$ and $\hat b$ are the annihilation operators 
 for the photon and magnon  mode with quantum number $n$ .
 The photon frequency and coupling are given by (\ref{Eq:DickeDispersion}) with 
  $\Omega_{Sw} = \Omega_{Sw} (q_n)$
 and $g=g(q_n)$. The last term in Eq.(\ref{Eq:DickeModelHamQuantized}) is the quantized 
 Zeeman interaction energy\cite{SupplMat} $d_F H_y M_y $. 
   
 The Hamiltonian Eq.(\ref{Eq:DickeModelHamQuantized}) is diagonalized in terms of the MP operators
  \begin{align} \label{Eq:MPcreationQuantized} 
  \hat d^\dagger_{j} = p_{j} \hat a^\dagger + 
   m_{j} \hat b^\dagger + \tilde p_{j} \hat a + 
   \tilde m_{j} \hat b
  \end{align}
 where $j= \{UP, LP\}$. Coefficients of mixing fractions can be chosen real and satisfy the normalization condition 
 $p_{j}^2 + m_{j}^2 - \tilde p_{j}^2 - \tilde m_{j}^2 =1$. 
 Their behaviour  is shown as function of $H_0$ in Fig.\ref{Fig:FractionsH0}a for the lower MP branch with $n=1$.
 The most exotic feature of MPs (\ref{Eq:MPcreationQuantized}) are squeezed vacuum states \cite{gerry2005introductory,klimov2009group}
  $| 0 \rangle_{sq}$ different separately for every quantized mode (\ref{Eq:MPcreationQuantized})  as $\hat d_{j} | 0 \rangle_{sq} = 0$. 
     Let us consider in more detail the vacuum states of the lower MP modes. Annihilation operators 
   for each $n$ can be written as $\hat d_{LP} = \hat S_m \hat S_p (\alpha \hat a + \beta \hat b) $ using magnon 
   $ \hat S_m = \exp [r_m (\hat b \hat b - \hat b^\dagger \hat b^\dagger)/2] $ and photon 
   $\hat S_p = \exp [r_p (\hat a \hat a - \hat a^\dagger \hat a^\dagger)/2]$ squeezing operators 
   \cite{gerry2005introductory,klimov2009group, SupplMat}.
   Here the squeezing parameters are $r_{p} = {\rm atanh} (\tilde p_{LP} /p_{LP} )$ and $r_m = {\rm atanh} (\tilde m_{LP} /m_{LP} )$.
   The vacuum state is obtained by squeezing both photons and magnons  
 $|0\rangle_{sq} = \hat S_p \hat S_m |0, 0\rangle$ where $|0, 0\rangle$ is the usual vacuum state. 
  As shown in Fig.\ref{Fig:FractionsH0}a squeeze parameters strongly depend on $H_0$. 
  For  $H_0 \ll M_0$ we get 
  $r_p= \Omega_{UP}/2\Omega_K$  
 and $r_m = (\Omega_{FMR}/\Omega_{Sw})\Omega_{UP}/2\Omega_K$, that is  divergence $r_{m,p} \sim \sqrt{M_0/H_0}$. 
 { Hence in the limit $H_0\to 0 $ such squeezing can become quite large compared to the highest known photon squeezing with \cite{schnabel2017squeezed, vahlbruch2016detection} 
 $r_p =1.7$ and even sublattice magnon squeezing $r_m \sim 3$ in antiferromagnets \cite{kamra2019antiferromagnetic}. }
 It can be used for quantum sensing applications 
 \cite{degen2017quantum}, generation of non-classical photon\cite{schnabel2017squeezed} and magnon states \cite{PhysRevLett.116.146601,PhysRevLett.122.187701} as well as of the previously unknown mutual photon-magnon entanglement. 
  
  Squeezed vacuum states are characterized by the non-zero density of virtual excitations\cite{ciuti2005quantum,de2017virtual,frisk2019ultrastrong}, which in our case 
  are magnons and photons.  The populations of virtual photons and magnons are
  $n_{p} = \langle 0|  \hat  a^\dagger \hat a |0 \rangle_{sq} = \sinh (2r_p)$ and 
  $n_{m} = \langle 0|   \hat b^\dagger \hat b|0 \rangle_{sq} = \sinh (2r_m)$, respectively. 
Our S/FI/S and S/FM/I/S system  host very large populations of virtual magnons and photons especially at $H_0\ll 4\pi M_0$
 where populations diverge exponentially as shown in Fig.\ref{Fig:FractionsH0}a.
  Such large populations of virtual bosons make our system very promising for the generation of entangled photon and magnon pairs through the analogue of dynamical Casimir effect \cite{frisk2019ultrastrong, de2017virtual, lahteenmaki2016coherence, lahteenmaki2013dynamical, 
moore1970quantum, nation2012colloquium, paraoanu2020listening,
wilson2011observation, johansson2010dynamicalA, johansson2009dynamical, ciuti2005quantum}. Note that one can easily achive abrupt the 
time-dependent vacuum state populations needed for the efficient generation photon and magnon pairs \cite{ciuti2005quantum} by varying $H_0$ or $\lambda$ faster than the MP frequencies $\Omega_{LP,n}$ .  
      
 Finally, let us demonstrate the bipartite entanglement between photons and magnons in the excited MP states. 
 Such type of entanglement has been not discussed before. 
 Acting with the lower  MP creation operator on the corresponding  vacuum state 
 we get the  excited state given by $ |1\rangle = \hat d^\dagger_{LP} |0\rangle_{sq}$. 
 This state consists \cite{SupplMat} of non-separable magnon and photon parts. 
 Their entanglement is determined by the von Neumann entropy  \cite{gerry2005introductory, nishioka2018entanglement}
 $ S_{mp} = -  {\rm Tr} (\hat \rho_p \ln \hat \rho_p )$. Here  
 the reduced density matrix is calculated taking the trace over magnon states $\hat\rho_p = 
 {\rm Tr}_m \langle \psi_m |\hat\rho|\psi_m \rangle $ from the full density matrix corresponding to the pure excited
 MP state is $\hat\rho = |1\rangle  \langle 1| $. The resulting dependencies of $ S_{mp} (H_0)$ for the lower MP modes with $n=1;\;2;\;3$
 is shown in Fig.\ref{Fig:FractionsH0}b for the S/FM/I/S system with parameters same as in Fig.\ref{Fig:DispersionMPH0} and experiment\cite{golovchanskiy2021approaching}. Such a high von Neumann entropy has been measured only in cold atomic systems\cite{islam2015measuring}.
 Potentially it can have many practical applications in transferring entanglement between different types of quantum systems. 
        
To summarize, we have found theoretically the mechanism of the ultrastrong magnon-photon coupling in superconducting nanostructures consisting of superconducting, ferromagnetic and insulating layers. This coupling leads to the highly squeezed vacuum states with 
large number of virtual photons and magnons at microwave frequencies. Our theory yields  magnon-polariton modes with  ultra-high cooperativity propagating with the velocity of about $10^4$ km/s. The calculated magnon-polariton frequency spectrum  explains  recent 
experiments very accurately. Excited magnon-polariton states are shown to consists of entangled magnon and photon states with large bipartite von Neumann entropy. These wonderful properties of magnon-polariton states put forward the suggested S/FI/S and S/FM/I/S systems as 
promising platforms for various magnonics applications.
   
  Many stimulating discussions with Igor Golovchanskiy, Vladmir Krasnov and Alexander Mel'nikov were very useful for this work . 
     
   \appendix
  
  \section{Derivation of dispersion relation Eqs.(1,5,6) 
  for S/FI/S structure}
  \label{Sec:Appendix1}

 Here we derive the dispersion relations for MP modes in 
 S/FI/S systems shown in Fig.1a. 
  First, let us consider the Maxwell equation 
 \begin{align}
  \label{EqApp:HomegaMaxwellWave}
 & \bm\nabla\times ( {\tilde \varepsilon}^{-1} \bm\nabla\times \bm H_\omega ) - q_v^2 \bm B_\omega =0
  \end{align} 
  %
  where $q_v = \omega/c$ and $\tilde{\varepsilon} = \varepsilon - 4\pi i \sigma/\omega$. 
  In components it yields   
   \begin{align} \label{EqApp1:MaxwX}
 & iq  \nabla_z H_z - \nabla_z^2 H_x  = \tilde{\varepsilon} q_v^2 H_x 
    \\  \label{EqApp1:MaxwZ}
  & iq  \nabla_z H_x + q^2 H_z  = \tilde{\varepsilon} q_v^2 B_z  
   \\ \label{EqApp1:MaxwY}
   &q^2 H_y - \nabla_z^2 H_y   = \tilde{\varepsilon} q_v^2 B_y  
   \end{align}
  In addition, from Landau-Lifshiz-Gilbert (LLG) equation we have the relation between the components of $B_{y,z}$ and 
  $H_{y,z}$ which in general yields  
  \begin{align} 
   & \left( B_y \atop B_z \right) = \begin{pmatrix}
   \gamma H_0 & -i \omega
   \\
   i \omega  & \gamma H_0
\end{pmatrix}^{-1}
 \begin{pmatrix}
  \gamma  B_0 & -i \omega
   \\
   i \omega & \gamma  B_0
\end{pmatrix}   
 \left( H_y \atop H_z \right)
     \end{align}
  In general all three equation in system (\ref{EqApp1:MaxwX},\ref{EqApp1:MaxwZ},\ref{EqApp1:MaxwY})
  are coupled to each other. However in the presence of metallic layers it is  significantly simplified
  because with good accuracy $B_z=0$. Indeed, in S we have $\tilde{\varepsilon} q_v^2 = - i \lambda^{-2} $
  and Eq.(\ref{EqApp1:MaxwZ}) $B_z \sim (q \lambda)^2 H_{x,z} $. 
   Hence, in the long-wave limit  $q \lambda \ll 1$ 
   we have $B_z \ll  H_{x,z} $ and with good accuracy can set $B_z=0$. 
  In this case the LLG equation reduces to the scalar relation  
  \begin{align} \label{EqApp1:ByHy}
  B_y= H_y(\Omega_B^2 -\omega^2)/(\Omega_K^2 -\omega^2) 
  \end{align}
  where $\Omega_B= \gamma B_0$ and $\Omega_K = \gamma \sqrt{H_0B_0}$.
  Therefore  Eq.\ref{EqApp1:MaxwY} for $H_y$ decouples from others and becomes 
    {
 \begin{align} 
  \label{EqApp1:HyS}
 {\rm in\; S:}\;\;\;\;\;\; & \nabla^2_z  H_y - \lambda^{-2}  H_y = 0
  \\ \label{EqApp1:HyFI}
 {\rm in\; FI:}\;\;\;\;\;\; & \nabla^2_z  H_y - \left(q^2- \varepsilon q_v^2 \frac{\omega^2-\Omega_B^2}{\omega^2-\Omega_K^2} \right)  H_y = 0
  \end{align}  
 Boundary conditions at interfaces  follow directly from Maxwell equations yielding  the continuity of tangential components $H_y$ and $E_x$ fields  
 \begin{align} \label{EqApp1:BCmi}
 [H_y]  = 0; \;  [E_x]  = 0
 \end{align}  
 where  $[...]$ denotes the jump  across the interface. The electric field is determined by the Faraday law 
 $i q_v \tilde \varepsilon \bm E = {\bm \nabla} \times \bm H$. It yields  
$E_x=(i/\tilde \varepsilon q_v)\nabla_z H_y  $ which is valid both  in metallic and insulating layers.  
 }


       Equation for $H_y$ is supplemented by the boundary conditions at the S/FI interface. They are the continuity of $H_y$ and $E_x$ components. The latter is given by  
   \begin{align}
   {\rm in\; S:}\;\; & E_x = -i q_v\lambda^2  \nabla_z H_y 
   \\
   {\rm in\; FI \;:}\;\; & E_x = (i/\varepsilon q_v)\nabla_z H_y
   \end{align}
  To get this relations we used that 
  $c/4\pi\sigma_F = i q_v l_{sk}^2 $ and 
  $c/4\pi\sigma_S = i q_v \lambda^2$. 
  
   We consider the S/FI/S system with S layers much thicker than $\lambda$. 
Solution is symmetric with respect to the middle of F layer at $z=0$:
 \begin{align} 
  \label{EqApp1:HyFsol}
 {\rm in\; FI:}\;\; & H_y= h_{FI} \cosh(q_z z);\;\;  
 \\
 & E_x=  i (q_z /\varepsilon q_v) h_{FI} \sinh(q_z z)
 \\ \nonumber
 \\ \label{EqApp1:HySsol}
 {\rm in\; S:}\;\; & H_y= h_S e^{-z/\lambda};\;\; 
 \\
 & E_x=   i q_v\lambda  h_S e^{- z/\lambda}
 \end{align}

  From the boundary conditions at FI/S interface $z=d_{F}/2$ 
  we get equation 
  \begin{align}\label{EqApp1:disp0}
  \tanh (q_z d_{F}/2) = \varepsilon \lambda q_v^2/q_z
  \end{align}
  In the long-wave limit $q d_{F} \ll 1 $ we use expansion 
  \begin{align}\label{EqApp1:disp1}
   (q_z/q_v)^2 d_{F}/2\varepsilon\lambda =1
  \end{align}
  where $q_z^2=q^2- \varepsilon q_v^2 \frac{\omega^2-\Omega_B^2}{\omega^2-\Omega_K^2}$.
  This equation can be rewritten as 
  \begin{align}
   \omega^4 - \omega^2 (\Omega_{Sw}^2 + \Omega_{FMR}^2 ) + 
   \Omega_K^2 \Omega_{Sw}^2 =0
  \end{align}
  and finally 
   \begin{align}\label{EqApp1:MPdispersion}
   (\omega^2 - \Omega_{Sw}^2)(\omega^2 - \Omega_{FMR}^2) = 
   (\Omega_{FMR}^2 - \Omega_K^2) \Omega_{Sw}^2
  \end{align}
  where 
  \begin{align}  \label{EqApp1:SwSFIS}
   & \Omega_{Sw} (q)   =  
 cq \sqrt{ d_{F}/\varepsilon( d_{F} + 2\lambda) }
   \\
  \label{EqApp1:OmegaFMRFI}
  &  \Omega_{FMR}  =  
 \sqrt{\Omega_K^2 + \frac{d_{F} \Omega_B\Omega_M }{  d_{F} + 2\lambda  }  }  
  \end{align}   
  These Eqs.(\ref{EqApp1:MPdispersion},\ref{EqApp1:SwSFIS},\ref{EqApp1:OmegaFMRFI})
  are the MP dispersion in S/FI/S system used in the main text.

  \section{Derivation of dispersion relation Eqs.(2,5,7) for S/FM/I/S structures}
  \label{Sec:Appendix3}

  Here we derive the dispersion relations for MP modes in 
 S/FM/I/S system shown in Fig.1b. 
 First, we write Maxwell equation for $H_y$
 Therefore  Eq.\ref{EqApp1:MaxwY} for $H_y$ decouples from others and becomes 
    {
 \begin{align} 
  \label{EqApp3:HyS}
 {\rm in\; S:}\;\;\;\;\;\; & \nabla^2_z  H_y - \lambda^{-2}  H_y = 0
  \\
  \label{EqApp3:HyFM}
 {\rm in\; FM:}\;\;\;\;\;\; & \nabla^2_z  H_y - l_F^{-2}  H_y = 0
  \\ \label{EqApp3:HyI}
 {\rm in\; I:}\;\;\;\;\;\; & \nabla^2_z  H_y - \left(q^2- \varepsilon q_v^2  \right)  H_y = 0
  \end{align} 
  where $l_F = l_{sk} \sqrt{(\Omega_K^2-\omega^2)/(\Omega_B^2-\omega^2)}$ and the skin length is
    $l_{sk}= \sqrt{c^2/4\pi i \omega\sigma_F} $. 
 } 
  
  Then, we note that Maxwell equation yields the following relations between electric 
   and magnetic field 
   \begin{align}
   {\rm in\; FM:}\;\; & E_x= - i q_vl_{sk}^2 \nabla_z H_y 
   \\
   {\rm in\; S:}\;\; & E_x = -i q_v\lambda^2  \nabla_z H_y 
   \\
   {\rm in\; I \; and\;  FI :}\;\; & E_x = (i/\varepsilon q_v)\nabla_z H_y
   \end{align}
  To get this relations we used that 
  $c/4\pi\sigma_F = i q_v l_{sk}^2 $ and 
  $c/4\pi\sigma_S = i q_v \lambda^2$, where $l_{sk}$ is the screening length in FM.  

  Let us write solution in different layers
 \begin{align} 
  \label{EqApp3:HyFsol}
   {\rm in\; outer \; S:}\;\; 
   & H_y= e^{\mp \tilde z/\lambda}
  ;\;\; 
   \\ \nonumber
  & E_x= \pm (iq_v \lambda) e^{\mp \tilde z/\lambda}
  \\ \nonumber
 \\ \label{EqApp3:HySsol}
 {\rm in\; FM:}\;\; 
 & H_y = h_{1F} e^{\tilde z/l_F} + h_{2F} e^{-\tilde z/l_F}
 \\
 & E_x= - \frac{i q_v l_{sk}^2}{l_F} 
 (h_{1F} e^{\tilde z/l_F} - h_{2F} e^{-\tilde z/l_F}) 
 \\ \nonumber  
  \\ \label{EqApp3:HyIsol}
 {\rm in\; I:}\;\; 
 &  H_y = h_{1I} e^{q_z \tilde z} + h_{2I} e^{-q_z\tilde z}
 \\ \nonumber
 & E_x= \frac{iq_z}{\varepsilon q_v} ( h_{1I} e^{q_z\tilde z} - 
 h_{2I} e^{-q_z\tilde z})
  \end{align}  
  where $q_z= \sqrt{q^2 - \varepsilon q_v^2 }$.
 Comparing the ratios $E_x/H_y$ across all the four boundaries we get   from the boundary conditions at different interfaces 
 \begin{align} 
  \label{EqApp3:bcIS}
  & {\rm I/S\; outer:}\;\; 
       \frac{h_{1I} - h_{2I}}{h_{1I} + h_{2I}}
     = - \alpha
  \\ \nonumber
 \\ \label{EqApp3:bcSI}
& {\rm FM/I:}\;\; 
   \frac{h_{1I}e^{q_z d_I} - h_{2I}e^{-q_z d_I}  }
 {h_{1I}e^{q_z d_I} + h_{2I}e^{-q_z d_I}} 
 = - \alpha
\frac{l_{sk}^2}{\lambda l_F} 
 \frac{h_{1F} - h_{2F}}{h_{1F} + h_{2F}}
 \\ \nonumber
  \\ \label{EqApp3:bcSF}
 & {\rm S\; outer/FM:}\;\; 
   \frac{ h_{1F}e^{d_F/l_F} - h_{2F}e^{-d_F/l_F }  }
 {h_{1F}e^{d_F/l_F} + h_{2F}e^{-d_F/l_F }} 
 = 
 - \frac{\lambda l_F}{l_{sk}^2}
  \end{align}  
  where  $\alpha = \varepsilon q_v^2 \lambda/q_z $.
  We start from Eq.\ref{EqApp3:bcSF} and expanding by small parameter 
  $d_F/l_F$ we get
  $$ 
  \frac{h_{1F} - h_{2F}}{h_{1F} + h_{2F}} = -\frac{\lambda}{l_F} 
  \left( \frac{d_F}{\lambda}  +  \frac{l_F^2}{l_{sk}^2} \right)
  $$
  Substituting this into remaining equations  yields
   \begin{align}
 &    \frac{h_{1I} e^{2q_zd_I} - h_{1I}}{h_{1I} e^{2q_zd_I} + h_{1I}} = \alpha (1 + \beta)
   \end{align}
   where $ \beta= (d_{F}/\lambda) (l_{sk}/l_F)^2 $. 
 Solving this system and expanding by small parameter $q_z d_I$
  we get 
  \begin{align}
   1+ \frac{d_{F}}{2\lambda} \frac{l_{sk}^2}{l_F^2}
   =  \frac{q_z^2}{q_v^2} \frac{d_I}{2\lambda}
\end{align}   
  Substituting expression for $q_z$ and collecting terms we get dispersion relation 
   \begin{align}\label{EqApp3:MPdispersion}
   (\omega^2 - \Omega_{Sw}^2)(\omega^2 - \Omega_{FMR}^2) = 
   (\Omega_{FMR}^2 - \Omega_K^2) \Omega_{Sw}^2
  \end{align}
  where 
  \begin{align}
     \label{EqApp3:SwSFMIS}
 & \Omega_{Sw} (q)   =  
 cq \sqrt{ d_I/\varepsilon(d_I + d_{F} +2\lambda ) }
   \\
\label{EqApp3:OmegaFMRFM}
 &  \Omega_{FMR}  =  
 \sqrt{\Omega_K^2  + \frac{d_{F}\Omega_B\Omega_M }{d_I +  d_{F} + 2\lambda }}
  \end{align}   
  These Eqs.(\ref{EqApp3:MPdispersion},\ref{EqApp3:SwSFMIS},\ref{EqApp3:OmegaFMRFM})
  are the MP dispersion in S/FM/I/S system used in the main text. 
  

    \section{Derivation of vacuum state}
  \label{Sec:Appendix4}
    
    Our goal is to find the vacuum state of the lower MP 
    \begin{align}
    \hat d_{LP} = p_{LP}  \hat a + \tilde m_{LP}  \hat b^\dagger + m_{LP}  \hat b +
     \tilde p_{LP}  \hat a^\dagger
    \end{align}
       This operator can written using squeezing operators 
    \begin{align}
    \hat d_{LP} =
    \hat S_p \hat S_m (\alpha \hat a + \beta\hat b ) \hat S_p^\dagger S_m^\dagger
    \end{align}
    where the squeeze operators are  
    \begin{align}
  &  \hat S_p = \exp [r_p (\hat a \hat a - \hat a^\dagger \hat a^\dagger)/2]
    \\
   &  \hat S_p = \exp [r_m (\hat b \hat b - \hat b^\dagger \hat b^\dagger)/2]
    \end{align} 
    Action of these operators  
   \begin{align}
   \hat S_p\hat S_m (\alpha \hat a + \beta\hat b ) 
   \hat S_m^\dagger\hat S_p^\dagger 
   = 
   \\ 
   \alpha ( u_p \hat a + v_p \hat a^\dagger  ) + 
   \beta ( u_m \hat b + v_m \hat b^\dagger  )
  \end{align}       
 where $u_{m,p} = \cosh r_{m,p}$, $v_{m,p} = \sinh r_{m,p}$.
 Hence the coefficients are 
 \begin{align}
 p_{LP} = \alpha \cosh r_{p} 
 \\
  \tilde p_{LP} = \alpha \sinh r_{p}
 \\  
 m_{LP}  = \beta \cosh r_{m} 
 \\
 \tilde m_{LP} =  \beta \sinh r_{m} 
  \end{align}
  so that the squeeze parameters are $ r_p = {\rm atanh} (\tilde p_{LP} /p_{LP} )$ and $ r_m = {\rm atanh} (\tilde m_{LP} /m_{LP} )$. 
  
  The vacuum state $\hat d_{LP} |0\rangle_{sq}$
 is then  related to the usual one $\hat a (\hat b) |0\rangle_{ab} =0$  as
 \begin{align}
 |0\rangle_{sq} = \hat S_p \hat S_m |0\rangle_{ab}
 \end{align}    
  The explicit form of vacuum state wave function consisting of the separable photon and magnon states 
  is given by
  \begin{align}
  & |0\rangle_{sq} = \psi_{p0} \psi_{m0} 
  \\
 & \psi_{p0} =  \sum_{n=0}^{\infty} P_n | 2n\rangle; \;\;
  P_n = \frac{(\tanh r_p/2)^n }{ \sqrt{\cosh r_p} } \frac{ \sqrt{(2n)!}}{n!} 
  \\ 
& \psi_{m0} =  \sum_{n=0}^{\infty} M_n | 2n\rangle ; \;\;
 M_n = \frac{(\tanh r_m/2)^n }{ \sqrt{\cosh r_m} } \frac{\sqrt{(2n)!}}{  n!} 
   \end{align}
   
   The excited state of MP is given by $ |1\rangle = \hat d^\dagger_{LP} |0\rangle_{sq}$ 
   \begin{align}
  & |1\rangle = (\psi_{p1}\psi_{m0} + \psi_{p0}\psi_{m1} )/(N_p + N_m)
   \\   
  & \psi_{p1} = \sum_{n=0}^{\infty} (  \sqrt{2n+1} P_n p  +  \sqrt{2n+2}  P_{n+1}\tilde p ) | 2n+1\rangle 
   \\
  & \psi_{m1} =  \sum_{n=0}^{\infty} (  \sqrt{2n+1} M_n m  +  \sqrt{2n+2} M_{n+1}\tilde m  ) | 2n+1\rangle 
   \end{align}
   with the normalization coefficients 
   \begin{align}
   N_p = \sqrt{\sum_{n=0}^{\infty} (  \sqrt{2n+1} P_n p  +  \sqrt{2n+2}  P_{n+1}\tilde p )^2 }
   \\
   N_m = \sqrt{\sum_{n=0}^{\infty} (  \sqrt{2n+1} M_n m  +  \sqrt{2n+2}  M_{n+1}\tilde m )^2 }
  \end{align}
  The density matrix corresponding to the pure excited MP state is $\hat\rho = |1\rangle  \langle 1| $. 
    The reduced density matrix is calculated taking the trace over magnon states $\hat\rho_p = 
    {\rm Tr}_m \langle \psi_m |\hat\rho|\psi_m \rangle $. The trace is calculated using the orthogonality 
  $\langle\psi_{m1}\psi_{m0} \rangle$=0 
  \begin{align}
  \hat \rho_p = \frac{1}{N_p + N_m} | \psi_{p1} \rangle \langle\psi_{p1} | + 
  \frac{N_m}{N_p + N_m} | \psi_{p0} \rangle \langle\psi_{p0}  |
  \end{align}
  Then the von Neumann entropy characterizing magnon-photon entanglement 
  is calculated using the standard definition $ S_{mp} = -  {\rm Tr} (\hat \rho_p \ln \hat \rho_p )$
   \begin{align}
    S_{mp} = - \frac{N_p}{N_p + N_m} \ln ( \frac{N_p}{N_p + N_m}  )
     - \frac{N_m}{N_p + N_m} \ln ( \frac{N_m}{N_p + N_m}  )
   \end{align}

 \bibliographystyle{apsrev4-2}
  \bibliography{refs2}

 \end{document}